# Deep Learning Coherent Diffractive Imaging

Dillan J. Chang[1,†], Colum M. O'Leary[1,†], Cong Su[2,3,4], Salman Kahn[2,3,4], Alex Zettl[2,3,4], Jim Ciston[5], Peter Ercius[5] and Jianwei Miao[1*]

[1]Department of Physics & Astronomy and California NanoSystems Institute, University of California, Los Angeles, CA 90095, USA.

[2]Department of Physics, University of California at Berkeley, Berkeley, CA 94720, USA.

[3]Materials Sciences Division, Lawrence Berkeley National Laboratory, Berkeley, CA 94720, USA

[4]Kavli Energy NanoSciences Institute at the University of California, Berkeley, CA 94720, USA.

[5]National Center for Electron Microscopy, Molecular Foundry, Lawrence Berkeley National Laboratory, Berkeley, CA 94720, USA.

**We report the development of deep learning coherent electron diffractive imaging at sub-ångström resolution using convolutional neural networks (CNNs) trained with only simulated data. We experimentally demonstrate this method by applying the trained CNNs to directly recover the phase images from electron diffraction patterns of twisted hexagonal boron nitride, monolayer graphene and a Au nanoparticle with comparable quality to those reconstructed by a conventional ptychographic method. Fourier ring correlation between the CNN and ptychographic images indicates the achievement of a spatial resolution in the range of 0.70 and 0.55 Å (depending on different resolution criteria). The ability to replace iterative algorithms with CNNs and perform real-time imaging from coherent diffraction patterns is expected to find broad applications in the physical and biological sciences.**

Coherent diffractive imaging (CDI), which replaces the physical lens of a microscope with a computational algorithm [1], is revolutionizing the imaging and microscopy field [2]. In particular, ptychography, a powerful scanning CDI method, takes advantage of overlapped illuminations in the sample plane as a constraint to simultaneously reconstruct the complex exit wave of the sample and the illumination function [3,4]. Although ptychography was proposed to extract the phase differences of overlapped diffraction spots for crystalline samples in 1969 [5], the modern version of ptychography utilizing iterative algorithms to recover the phases of non-crystalline objects was demonstrated with coherent x-rays in 2007 [3], which was based on the original CDI experiment in 1999 [1]. Ptychographic CDI with iterative algorithms has found wide applications with synchrotron radiation, high harmonic generation, electron and optical microscopy [2,6-24]. Albeit powerful, iterative algorithms are not only computationally expensive, but also require practitioners to get algorithmic training to optimize the parameters and obtain satisfactory results. These difficulties have thus far prevented ptychography from being accessible to an even broader user community. One approach to overcome these difficulties is the replacement of iterative algorithms with deep learning using convolutional neural networks (CNN) [25-29]. CNNs have recently been demonstrated for CDI experiments, such as near-field phase retrieval in an optical setup [30] and the reconstruction of complex wave function from coherent x-ray diffraction patterns [31-32]. However, current experimental realizations of deep learning CDI require a large database from which to train the CNNs – numerous experiments with a specific physical setup must be performed first before any future predictions can be made on that same setup. Deep learning CDI would be a more general and powerful method if it became possible to perform real-time phase retrieval without the need for large data bases collected via experimentation. In this Letter, we report that such an approach is not only possible, but powerful enough to compete against conventional ptychographic reconstruction algorithms in atomic-resolution electron imaging. Three experiments with different electron microscopes and detectors on different samples presented here demonstrate the universality and potential of deep learning CDI for real-time, atomic-scale imaging.

Deep learning CDI with augmented data begins with the forward propagation of a coherent source. Figure 1(a) shows a diagram of a typical electron ptychography setup where a focused coherent source illuminates an object. The resulting wave function is propagated to the Fraunhofer regime and only the square of the amplitude of the wave function is measured by a pixel array detector. Mathematically, this forward process relates the object function and the measurement by

$$M(\mathbf{k}) = |\mathcal{F}[P(\mathbf{r}) \cdot O(\mathbf{r})]|, \qquad (1)$$

where $M(\mathbf{k})$ is the amplitude of the Fourier transform ($\mathcal{F}$), $P(\mathbf{r})$ and $O(\mathbf{r})$ are the complex probe and object functions, respectively. Since the phase of the Fourier transform is lost during measurement, the inverse of this forward process is nonlinear. Figure 1(b) shows the process of converting a random stock image found on the internet into a pure phase object, which is illuminated by a probe function to produce an exit wave. The amplitude and phase of the probe function is estimated based on the defocus and aberration of the electron optics. The diffraction intensity of the exit wave with Poisson noise is calculated to satisfy the oversampling requirement for phase retrieval [33]. The square root of the noisy diffraction intensity is used to train CNNs with an L1-norm loss function to recover the phases in the illuminated area (named a phase patch) [Fig. 1(c)]. In our experience, using randomly generated stock photos from the internet provides a rich source of entropy within the images to sufficiently train the CNNs without imposing any regularizations [34]. More examples of phase image generation, forward process and CNN performance as validation data can be found in Supplemental Fig. 1.

The architecture of the CNN is an encoder-decoder architecture [Fig. 1(d)], more commonly known as U-net [35], with skip connections between corresponding tensor sizes to prevent vanishing gradient issues. Detailed schematics of the residual layers can be found in Supplemental Fig. 2, with skip connections acting as concatenations that provide direct throughput within the architecture. As the diffraction intensity is heavily corrupted with Poisson noise, a low learning rate of $1.0\times10^{-4}$ and a high dropout rate of 0.2 is applied at every layer to prevent overfitting of noise while training with stochastic gradient descent. The trained CNN is used to directly map from the amplitudes of the Fourier transform to phase patches without any iteration. Next, a stitching method is developed to tile the phase patches to form a phase image [Fig. 1(e)]. As the zero frequency of the phase is irrecoverable from the CNN, two adjacent recovered phases differ in the overlapped region by an overall phase shift. The stitching method is therefore executed by minimizing an error function ($E$),

$$E = \sum_{i,j\neq i}\left[\mu(\phi_i \cdot S_{i,j}) - \mu(\phi_j \cdot S_{j,i})\right]^2, \qquad (2)$$

where $\mu$ returns the mean value, $\phi_i$ is the $i^{\text{th}}$ phase, and $S_{i,j}$ is a mask where the $i^{\text{th}}$ phase is overlapped with the $j^{\text{th}}$ phase. The error is minimized by gradient descent and a well-stitched phase image usually requires only a few iterations.

To demonstrate the versatility of deep learning CDI, we performed electron ptychography experiments on different samples using different microscopes. The first sample consists of two 5-nm-thick hexagonal boron nitride (hBN) flakes with a twisted interface [36]. The experiment was conducted on the TEAM I double-corrected S/TEM instrument at the National Center for Electron Microscopy, Molecular Foundry (electron energy: 300 keV, convergence semi-angle: 17.1 mrad, dose: $4.5\times10^6$ e Å$^{-2}$, dwell time: 0.87 ms and probe step size: 0.25 Å). The microscope was equipped with a Gatan K3 pixelated detector, which operated in electron-counting mode and was binned (×2) and windowed (×2) to 512×512 pixels. The diffraction patterns were further binned to 32×32 pixels in post-processing. Figure 2(a) shows a representative diffraction pattern with the presence of heavy noise. Conventional STEM imaging modes such as annular dark-field (ADF) imaging produce images with a poor signal-to-noise ratio [Fig. 2(b)]. To train the CNN, a probe function was analytically generated by parameterizing the aberration function up to second order with seven total parameters: one from defocus, two from twofold astigmatism, two from coma, and two from threefold astigmatism [37]. After completing training with 250,000 simulated diffraction patterns of stock images from the internet (Supplemental Fig. 2), the CNN was used to independently retrieve the individual phase patches from the experimental data, which were then stitched together to form a phase image. Figure 2(c) and (d) show the phase image and a magnified view of the twisted hBN sample, respectively. As a comparison, a phase reconstruction by the extended ptychographic iterative engine (ePIE) [38] and a magnified view are shown in Fig. 2(e) and (f), respectively. Fourier ring correlation (FRC) analysis between the CNN and ePIE reconstructions shows a good agreement between the two methods [Fig. 2(g)]. The slight difference at low spatial frequencies of the FRC curve is likely because the phase patches from the CNN were more uniformly stitched together than those reconstructed by ePIE. Based on the cutoff criteria of FRC = 0.5 and 0.143 [39], the resolution of the CNN phase image was quantified to be 0.71 and 0.53 Å, respectively [red and blue dashed lines in Fig. 2(g)], demonstrating that both methods consistently reconstructed the diffraction signal beyond the bright-field disk.

Next, we investigated the tolerance of the CNN phase retrieval to data sparsity by imaging monolayer graphene with varying overlap of the diffraction patterns. The ptychographic data set was acquired using a JEOL 4DCanvas pixelated detector installed on a JEM-ARM200F probe-corrected microscope (electron energy: 80 keV, convergence semi-angle: 31.5 mrad, dose: $1.4\times10^6$ e Å$^{-2}$, and dwell time: 0.25 ms), as reported elsewhere [40]. The probe function was analytically generated with a second order aberration function as implemented in the hBN experiment, and a CNN was trained by 250,000 simulated diffraction patterns (Supplementary Fig. 2). Phase patches independently retrieved by the CNN from the experimental diffraction patterns were stitched together to generate a phase image [Fig. 3(a)], which is in a good agreement with an ePIE reconstruction from the same data set [Fig. 3(e)]. To investigate the performance of the CNN under varying data sparsity conditions, every other diffraction pattern in both x and y scanning directions was used to reconstruct the phase image in Fig. 3(b), doubling the scanning step size from 0.132 Å to 0.264 Å and quartering the effective electron dose. Additionally, phase images of the CNN obtained by taking every three (0.396 Å step size) and four (0.528 Å step size) diffraction patterns are shown in Fig. 3(c) and (d), respectively. Note that the phase images in Fig. 3(a-d) were recovered using the same trained CNN, as the forward process did not change when taking data with varying scanning step sizes. As a comparison, the corresponding phase images reconstructed by ePIE are shown in Fig. 3(e-h), demonstrating that ePIE and CNN provide similar reconstruction qualities with varying overlap of the diffraction patterns.

Finally, a ptychographic experiment was conducted on a 5 nm Au nanoparticle to examine the performance of CNNs with strongly scattering atoms. The data set was acquired using the 4D Camera (576×576 pixels) [41] installed on the TEAM 0.5 double-corrected S/TEM instrument at NCEM (electron energy: 300 keV, convergence semi-angle: 17.1 mrad, dose: $4.6 \times 10^4$ e Å$^{-2}$, dwell time: 0.044 ms, and probe step size: 0.25 Å). Figure 4(a) shows a representative diffraction pattern after binning, where heavy noise is visible. An ADF-STEM image was generated by integrating the intensity outside the bright-field disk, which corresponds to the amplitude image of the diffraction patterns. The probe function was analytically generated with a second order aberration function as for the hBN and graphene cases, and a CNN was trained by 250,000 simulated diffraction patterns (Supplementary Fig. 2). Figure 4(c) and (d) show the phase images retrieved by the CNN and ePIE, respectively, exhibiting consistent atomic features along the zone axis. The FRC curve of the two images indicates a resolution of 0.70 and 0.55 Å with the cutoff criteria of FRC = 0.5 and 0.143, respectively [Fig. 4(e)]. The dip in the low spatial frequencies in the FRC curve is consistent with the observation in the hBN result [Fig. 2(g)].

We demonstrated deep learning CDI using CNNs trained only with stock images from the internet. The CNNs were then used to directly retrieve the phase images of monolayer graphene, twisted hBN and a Au nanoparticle from experimental electron diffraction patterns. Quantitative analysis using FRC curves indicates that the phase images recovered by the CNNs have comparable quality to those reconstructed by ePIE. The resolution of the phase images by the CNNs was quantified by the FRC to be in the range of 0.71-0.53 Å. Compared with ptychography that uses iterative algorithms and overlapped regions to reconstruct the phase information [3,4,38], deep learning CDI independently recovers phase patches at different scanning positions by CNNs and subsequently stitches them together to form a phase image, which can thus be implemented in real time. Looking forward, deep learning CDI could be combined with atomic electron tomography [42,43] to determine the 3D atomic structure of radiation sensitive, low-Z and amorphous materials [44-46]. Furthermore, although we focused on the recovery of phase objects in this work, the CNNs can in principle be extended to retrieve the complex exit wave of the sample [31,32]. The probe function could potentially be simultaneously recovered by CNNs from diffraction patterns without the accurate knowledge of the defocus and aberration of the electron optics [4,18,20,22]. With these improvements in mind, we expect that deep learning CDI could become an important tool for real-time, atomic-scale imaging of a wide range of samples across different disciplines.

This work was primarily supported by the US Department of Energy, Office of Science, Basic Energy Sciences, Division of Materials Sciences and Engineering under award no. DE-SC0010378. We also thank the support by STROBE: a National Science Foundation Science and Technology Center under award no. DMR-1548924 and the Army Research Office Multidisciplinary University Research Initiative (MURI) program under grant no. W911NF-18-1-0431. The electron ptychography experiments with TEAM I and TEAM 0.5 were performed at the Molecular Foundry, which is supported by the Office of Science, Office of Basic Energy Sciences of the US Department of Energy under contract no. DE-AC02-05CH11231. C.S. and A.Z. acknowledge the support by the Director, Office of Science, Office of Basic Energy Sciences, Materials Sciences and Engineering Division, of the US Department of Energy under contract no. DEAC02-05-CH11231, within the sp2 -Bonded Materials Program (KC-2207) which provided for preliminary TEM and Raman characterization of the h-BN material, and by the van der Waals Heterostructures program (KCWF16) which provided for assembly of the twisted h-BN material.

[†]These authors contributed equally to this work.

[*]miao@physics.ucla.edu

**Figures and Figure Captions**

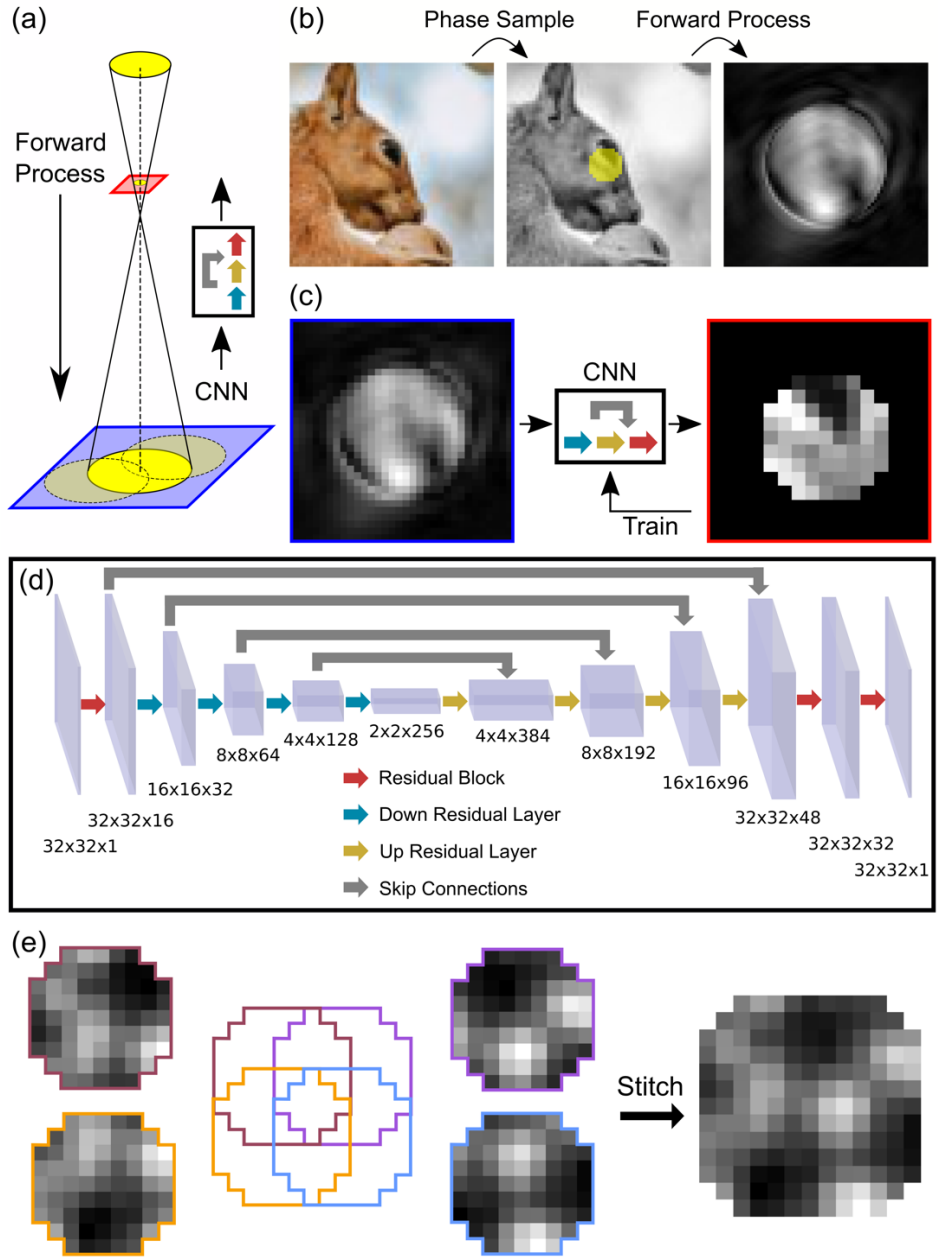

**Fig. 1.** Principle of deep learning electron diffractive imaging. (a) The forward process in electron ptychography that maps from an object to overlapped diffraction patterns. A CNN is trained to directly invert this forward process. (b) A random image is used to generate a phase sample, which is then used to calculate diffraction patterns using the forward process from (a), where the yellow circle indicates the illuminated area of one scanning position. (c) The CNN is trained to directly map from the square root of the diffraction patterns to the phase patches within the illuminated area. Note that each phase patch is treated fully independently in the CNN training and the overlapping regions of diffraction patterns are only used to align the recovered phase patches. (d) Detailed schematic of the CNN architecture. (e) Four representative phase patches directly retrieved by the CNN. Based on the overlapped regions, all the phase patches are stitched together to form a phase image.

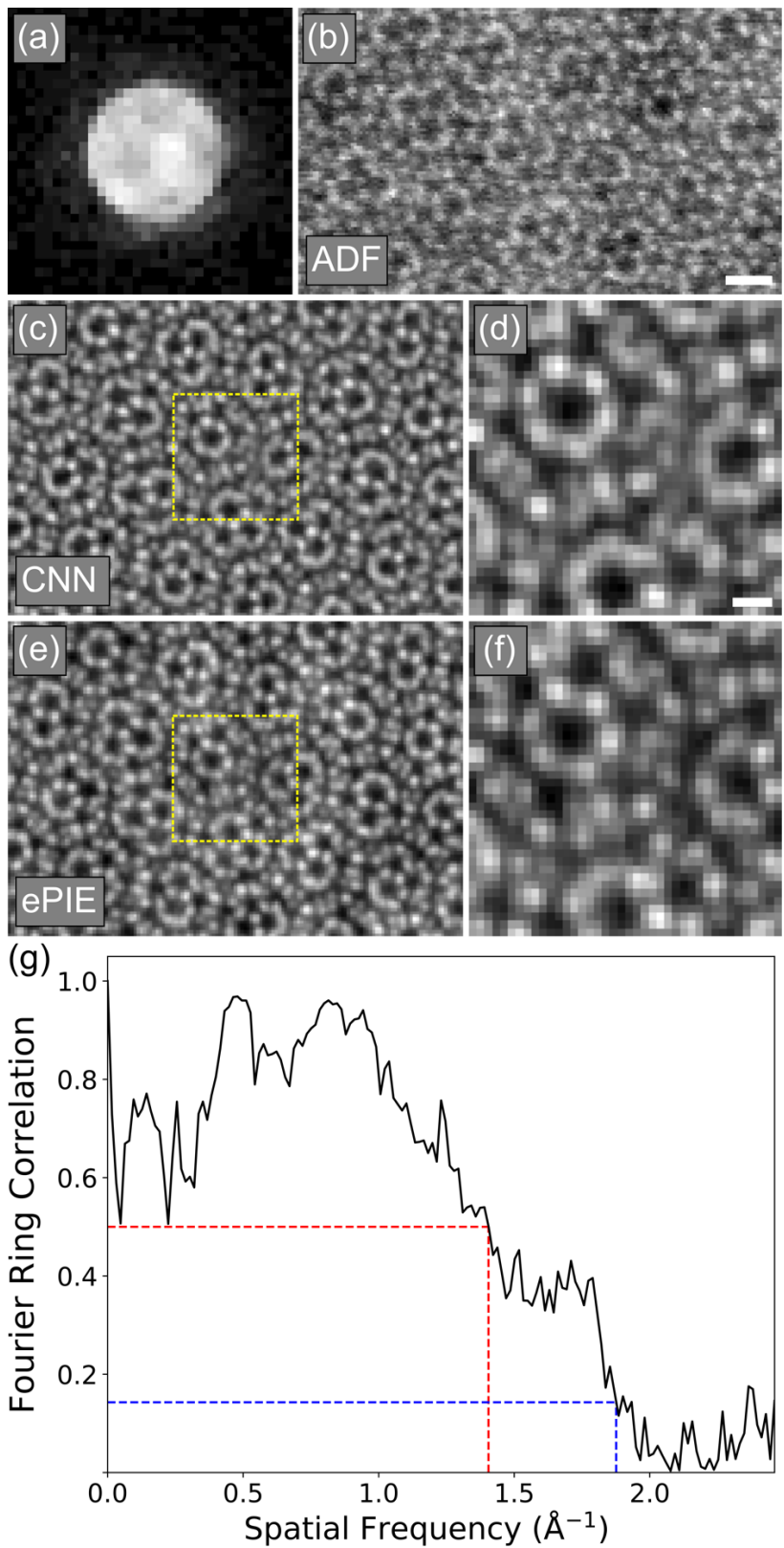

**Fig. 2.** Deep learning phase retrieval of an experimental data set of a twisted hBN interface. (a) Representative diffraction pattern measured from two 5-nm-thick hBN flakes with a twisted interface. (b) ADF-STEM image of the hBN sample generated by integrating the diffraction intensity outside the bright-field disk at each scanning position. (c) Phase image of the twisted hBN interface retrieved by a trained CNN. (d) Magnified view of the dotted square in (c). (e) Phase reconstruction of the twisted hBN interface reconstructed by ePIE. (f) Magnified view of the dotted square in (e). (g) Fourier ring correlation between (c) and (e), where the red and blue dashed lines indicate a resolution of 0.71 and 0.53 Å, based on the cutoff criteria of FRC = 0.5 and 0.143, respectively. Scale bars, 4 Å (b); and 2 Å (d).

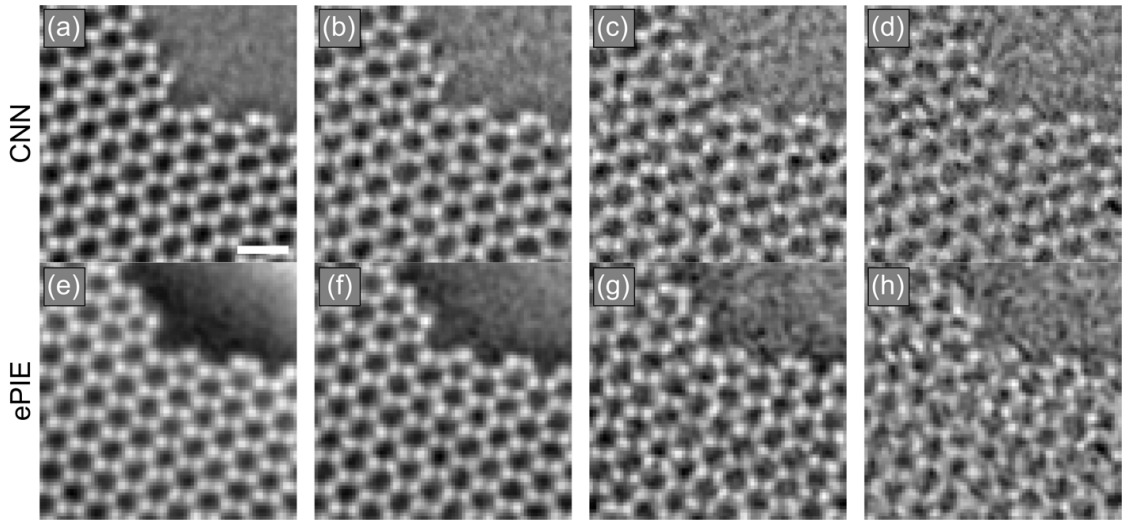

**Fig. 3.** Deep learning phase retrieval of an experimental graphene data set with varying the overlap of diffraction patterns. (a-d) Phase images of monolayer graphene retrieved by a trained CNN with a scanning step size of 0.132 Å, 0.264 Å, 0.396 Å, and 0.528 Å, respectively. (e-h) Phase images of the graphene sample reconstructed by ePIE corresponding to the same scanning step sizes in (a-d). Scale bar, 4 Å.

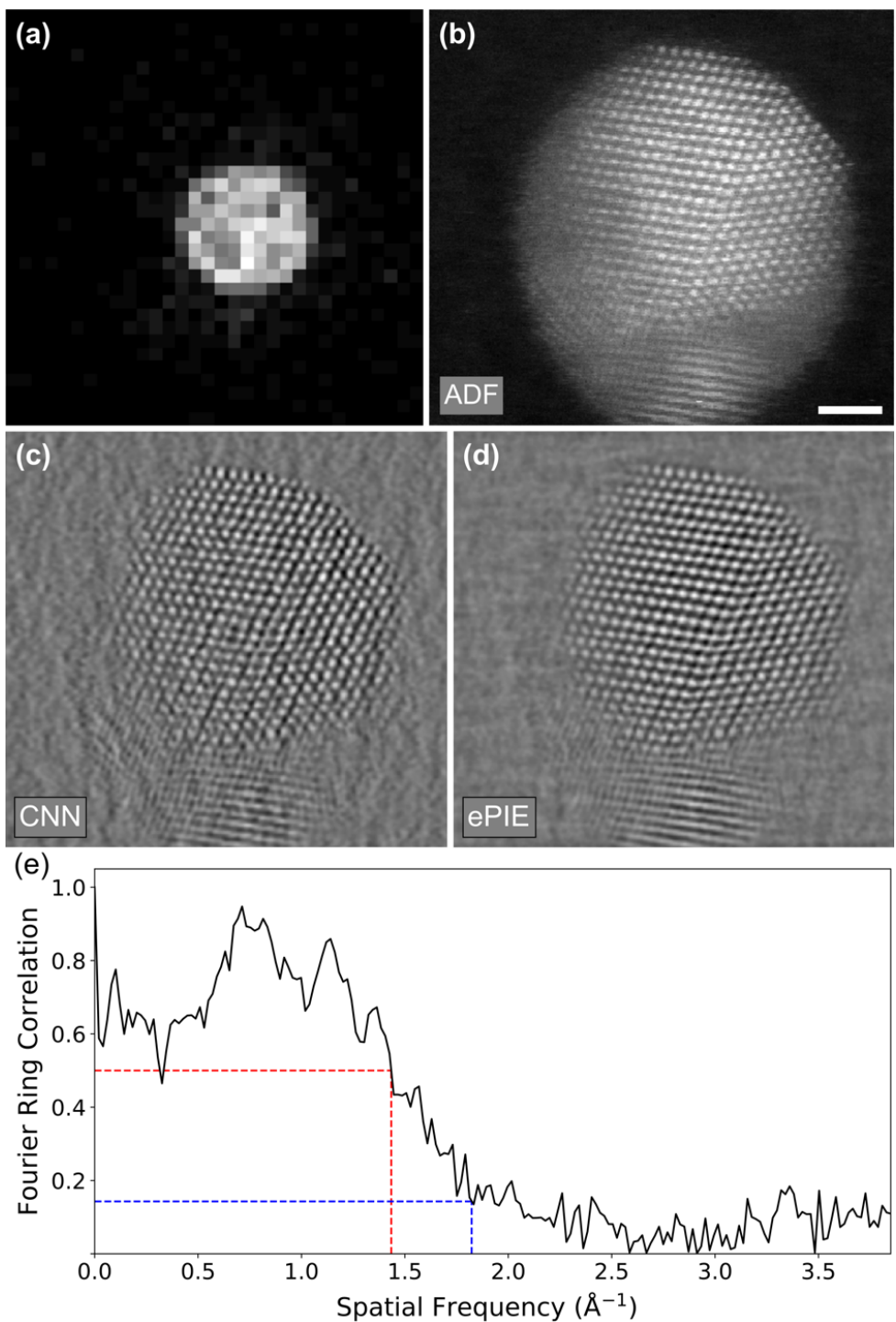

**Fig. 4.** Deep learning of phase retrieval of an experimental Au nanoparticle data set. (a) Representative diffraction pattern acquired from a 5-nm Au nanoparticle. (b) ADF-STEM image

generated by integrating the diffraction intensity outside the bright-field disk at each scanning position. (c) Phase image of the Au nanoparticle by a trained CNN. (d) Phase image of the nanoparticle by ePIE. (e) FRC curve between (c) and (d), where the red and blue dashed lines indicate a resolution of 0.70 and 0.55 Å, based on the cutoff criteria of FRC = 0.5 and 0.143, respectively. Scale bar, 1 nm.

# Supplementary Figures

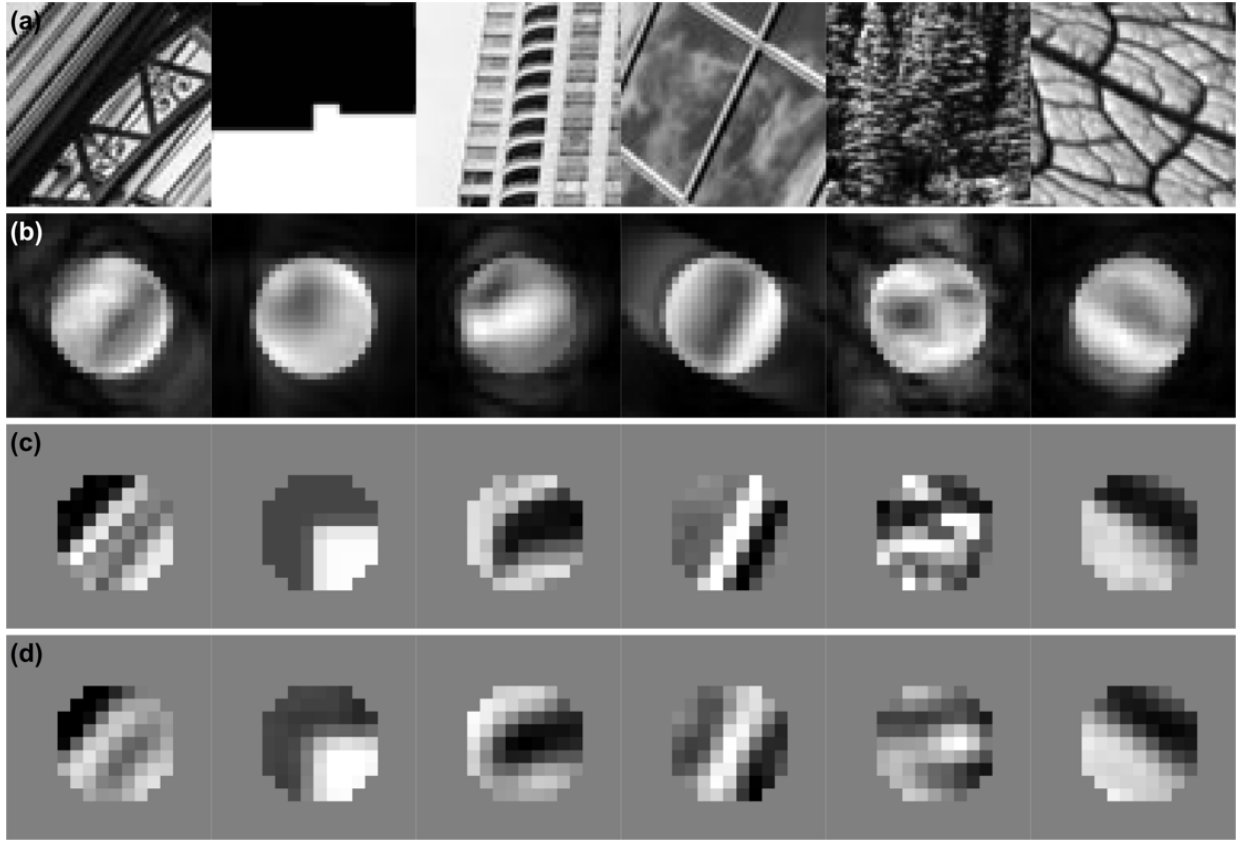

**Supplemental Fig. 1.** Examples of augmented data generation and CNN performance. (a) Random images from the internet used as pure phase objects to train a CNN. (b) Diffraction patterns generated from the phase objects in (a). (c) Perfect phase patches within the illuminated areas. (d) Corresponding phase patches independently retrieved from the square root of the diffraction patterns by a trained CNN without any iteration.

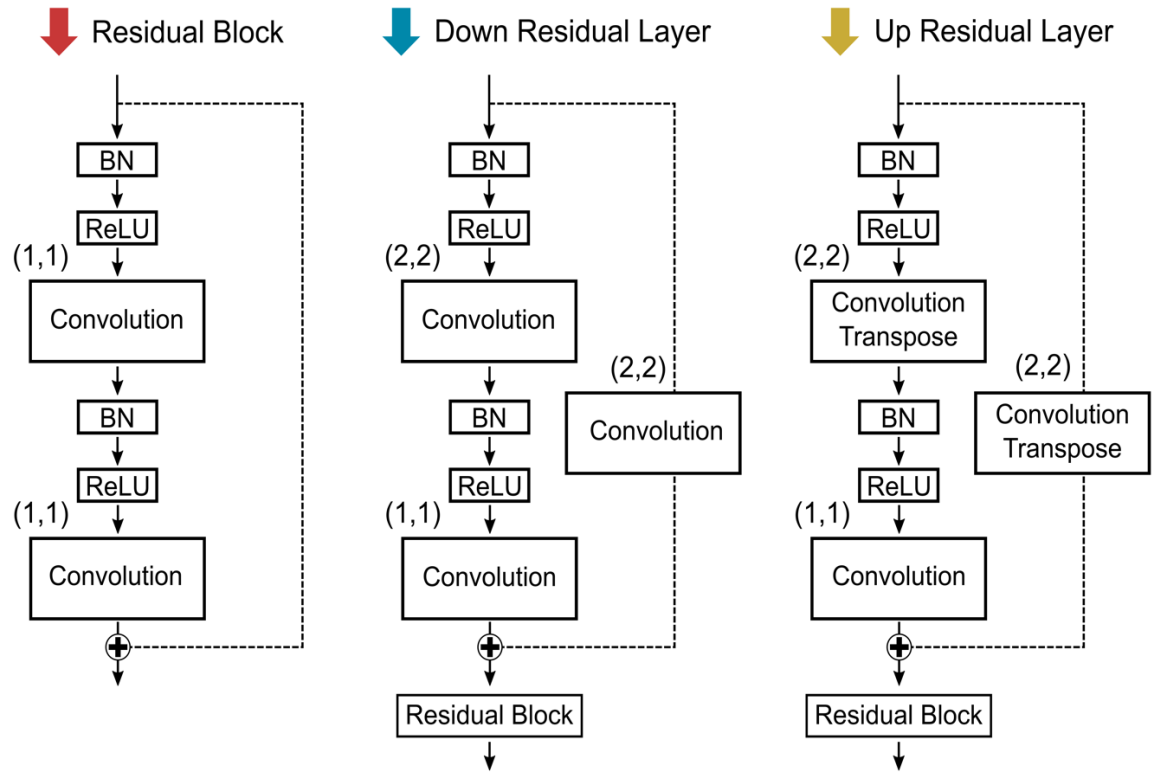

**Supplemental Fig. 2.** Schematic of residual layers used in the CNN architecture. Different strides used in the convolution filters are shown in parentheses. All filters are size 3×3.